\documentclass[%
 aip,
 amsmath,amssymb,
reprint,%
]{revtex4-1}

\usepackage{graphicx}
\usepackage{dcolumn}
\usepackage{bm}

\usepackage[utf8]{inputenc}
\usepackage[T1]{fontenc}
\usepackage{mathptmx}
\usepackage{etoolbox}
\usepackage{xcolor}

\makeatletter
\def\@email#1#2{%
 \endgroup
 \patchcmd{\titleblock@produce}
  {\frontmatter@RRAPformat}
  {\frontmatter@RRAPformat{\produce@RRAP{*#1\href{mailto:#2}{#2}}}\frontmatter@RRAPformat}
  {}{}
}%
\makeatother

\newcommand{\beq}{\begin{equation}}
\newcommand{\eeq}{\end{equation}}
\newcommand{\barr}{\begin{eqnarray}}
\newcommand{\earr}{\end{eqnarray}}
\newcommand{\bseq}{\begin{subequations}}
\newcommand{\eseq}{\end{subequations}}
\newcommand{\vett}[1]{\textbf{#1}}

\begin{document}

\title{Field theory description of the non-perturbative optical nonlinearity of epsilon-near-zero media}

\author{Yaraslau Tamashevich}
\affiliation{Faculty of Engineering and Natural Sciences, Tampere University,Finland}

\author{Tornike Shubitidze}
\affiliation{Department of Electrical \& Computer Engineering, Boston University, 8 Saint Mary's Street, Boston, 02215, MA, USA}

\author{Luca Dal Negro$^{*}$}\email{dalnegro@bu.edu}
\affiliation{Department of Electrical \& Computer Engineering, Boston University, 8 Saint Mary's Street, Boston, 02215, MA, USA}
\affiliation{Department of Physics, Boston University, 
590 Commonwealth Avenue, Boston,02215, MA, USA}
\affiliation{Division of Materials Science \&  Engineering, Boston University, 15 St. Mary’s street, Brookline, 02446, MA, USA}

\author{Marco Ornigotti$^{*}$}\email{marco.ornigotti@tuni.fi}
\affiliation{Faculty of Engineering and Natural Sciences, Tampere University,Finland}

\date{\today}

\begin{abstract}
In this paper we introduce a fully non-perturbative approach for the description of the optical nonlinearity of epsilon-near-zero (ENZ) media. In particular, based on the rigorous Feynman path integral method, we develop a dressed Lagrangian field theory for light-matter interactions and discuss its application to dispersive Kerr-like media with order-of-unity light-induced refractive index variations. Specifically, considering the relevant case of Indium Tin Oxide (ITO) nonlinearities, we address the novel regime of non-perturbative refractive index variations in ENZ media and establish that it follows naturally from a scalar field theory with a Born-Infeld (BI) Lagrangian. Moreover, we developed a predctive model that includes the intrinsic saturation effects originating from the light-induced modification of the Drude terms in the linear dispersion of ITO materials. Our results extend the Huttner-Barnett-Bechler electrodynamics model to the case of non-perturbative optical Kerr-like media providing an intrinsically nonlinear, field-theoretic framework for understanding the exceptional nonlinearity of ITO materials beyond traditional perturbation theory.
\end{abstract}


\maketitle

\section{Introduction}\label{sec1}
\noindent Epsilon-near-zero (ENZ) low-index optical materials exhibit a number of fascinating optical properties including a dramatic enhancement of nonlinear processes such as harmonic generation, frequency mixing and conversion, electro-optical modulation, and light-induced refractive index changes \cite{Liberal2017,Reshef2019_ENZ_Review, Niu2018, Wu2021, Khurgin2021, Ciattoni2016, Guo2016, Bohn2021, Vezzoli2018_Time_Reversal,Capretti2015THG}. These effects are boosted in sub-wavelength nanostructures across their ENZ spectral regions and  provide unprecedented opportunities for classical and quantum
information technologies, optical spectroscopy, and sensing. Specifically, indium tin oxide (ITO) degenerate semiconductors  have recently attracted a significant interest due to their extreme optical nonlinearity resulting in order-of-unity light-induced refractive index changes with sub-picosecond response time  \cite{Alam2016,Khurgin2021}. A number of impressive ENZ-driven optical nonlinear
phenomena and devices have been demonstrated based on ITO
nanolayers, including enhanced second-harmonic \cite{Capretti2015SHG} and third-harmonic
frequency generation \cite{Capretti2015THG}, ultrafast all-optical modulation \cite{Alam2016},
high-efficiency optical time reversal \cite{Vezzoli2018_Time_Reversal}, as well as  time-varying and spatio-temporal photonic devices \cite{Double_slit,Guo2016,Bohn2021,Gosciniak2023,Tirole2022_timeMirror}. 
Moreover, the linear optical dispersion properties of ENZ materials and structures based on ITO can be tuned widely with a zero permittivity
wavelength $\lambda_{ENZ}$ (i.e., the wavelength at which the real part of the permittivity vanishes) that extends from the near-infrared to mid-infrared spectral range \cite{Wang2015,Gui2019}, enabling the engineering of efficient nonlinear phenomena across multiple wavelengths. 
It was also established that a significant enhancement of the nonlinear response of ITO nanolayers can be achieved through
control of the material microstructure, specifically by exploiting secondary phases with anisotropic texturing that are largely influenced by the fabrication conditions \cite{Britton2022}.
Recently, Reshef et al. \cite{Reshef2017} observed that the conventional expression of the intensity-dependent refractive index loses its meaning for nonlinear ITO materials in the ENZ regime because the refractive index cannot be expanded in a perturbation series. In particular, they found that the contributions of susceptibility terms up to seventh-order to refraction are much bigger than the linear term and, consequently, the nonlinear response of ENZ and low-index media can no longer be interpreted as a small perturbation to linear optics.
However, as the authors clearly pointed out, the power series expansion of the nonlinear polarization in ENZ materials does not diverge at the ENZ wavelength. 
The intrinsically non-perturbative response of ITO thin films has been further enhanced by Shubitidze et al. \cite{Tornike23} through the coupling with surface-localized plasmon-polariton Tamm states in nonlinear photonic structures.  

\noindent In this work, we consider the nonlinear optical response of ENZ media from an alternative theoretical perspective and we develop, based on the rigorous Feynman path integral approach, a novel framework for understanding non-perturbative nonlinear refraction effects, such as the ones recently demonstrated in ITO low-index media, based on nonlinear field theory. In particular, our theory shows how non-perturbative refractive responses emerge naturally in dispersive Kerr-like media from an intrinsically nonlinear field theory characterized by the Born-Infeld (BI) Lagrangian. While focused for simplicity on a scalar case, our approach introduces a novel framework, more general than traditional nonlinear optics, for gaining a more fundamental understanding of the exceptional nonlinearity of ENZ materials beyond perturbation theory.

\noindent Our work is organised as follows: in section \ref{ENZintro} we briefly introduce the non-perturbative nonlinear refraction of ENZ media within a phenomenological model based on the concept of ``field-corrected" permittivity and illustrate the significant impact of low-index on the nonlinear response of ITO thin-films in the ENZ regime. In section \ref{theory1} we give a brief introduction to the concept and physical meaning of Feynman path integrals and contextualise their use in optics. We also introduce the basic model, i.e., the Huttner-Barnett-Bechler Lagrangian for light-matter interaction in dispersive ENZ media at the microscopic level, and derive the effective dressed electromagnetic Lagrangian, which constitutes the starting point of this work.

In section \ref{theory3} we introduce the nonlinearity and discuss the usual perturbative case, typical of nonlinear optics while in section  \ref{section5} we propose a new approach, based on the Born-Infeld Lagrangian, that goes beyond the perturbative limit of ENZ media and includes the distinctive saturation effects resulting from the bleaching of the Drude linear dispersion. Finally, in section \ref{conclusions} we draw our conclusions and future perspective.

\section{Nonlinear optical response of ENZ media} \label{ENZintro}
\noindent The objective of this section is to review the standard perturbative treatment of the induced nonlinear polarization of a Kerr-like medium with intensity-dependent refractive index and to introduce the necessary modifications required to describe ENZ low-index materials. We assume for simplicity a centrosymmetric material (i.e., scalar susceptibility) and neglect its magnetic response. Under these assumptions, the total polarization induced by monochromatic light can be written as: 
\begin{eqnarray}
P^{tot}(\omega)=P^{(1)}(\omega)+P^{(3)}(\omega)+\ldots=\epsilon_{0}\Big[\chi^{(1)}(\omega)+  \\ \nonumber
3\chi^{(3)}(\omega)|E(\omega)|^{2}+\ldots\Big]E(\omega),
\end{eqnarray}
where $E(\omega)$ is the incident electric field, $\chi^{(1)}$ is the linear susceptibility of the medium, and $\chi^{(3)}$ is its lowest-order nonlinear susceptibility. From the above expression 
we introduce the ``field-corrected" susceptibility \cite{butcher1990elements}:
\begin{equation}
	\chi_{fc}(\omega)\equiv
	\chi^{(1)}(\omega)+3\chi^{(3)}(\omega)|E(\omega)|^{2}+\ldots,
\end{equation}
and the ``field-corrected" dielectric function:
\begin{equation}
	\epsilon_{fc}(\omega)=1+\chi_{fc}(\omega).
\end{equation}
The linear refractive index is $n_{0}(\omega)=\sqrt{1+\chi^{(1)}(\omega)}
=\sqrt{\epsilon_{r}^{(1)}(\omega)}$ and $\epsilon_{r}^{(1)}$ is the linear contribution to the relative permittivity. In conventional nonlinear materials, susceptibility terms beyond the third-order are negligible in magnitude and the light induced refractive index change $\Delta{n}=n-n_{0}$ is very small under the usual assumption $\Delta{n}<<n_{0}$. Therefore, in these cases the change in  refractive index induced by the incident field can be well-approximated as follows:
\begin{equation}
	\Delta{n}(\omega)=\sqrt{\epsilon_{fc}(\omega)}
	-n_{0}(\omega)\approx\frac{3\chi^{(3)}|E|^{2}}{2n_{0}(\omega)},
\end{equation}
where we made use of the binomial expansion. The expression above leads to the conventional definition of the intensity dependent refractive index:
\begin{equation}\label{nlindex}
n=n_{0}+n_{2}I,
\end{equation}
where:
\begin{equation}\label{eqn6}
	n_{2}=\frac{3\chi^{(3)}}{4\epsilon_{0}{c}n^{2}_{0}},
\end{equation}
is the nonlinear refractive index of the medium.
\begin{figure}
\begin{center}
\includegraphics[width=0.4 \textwidth]{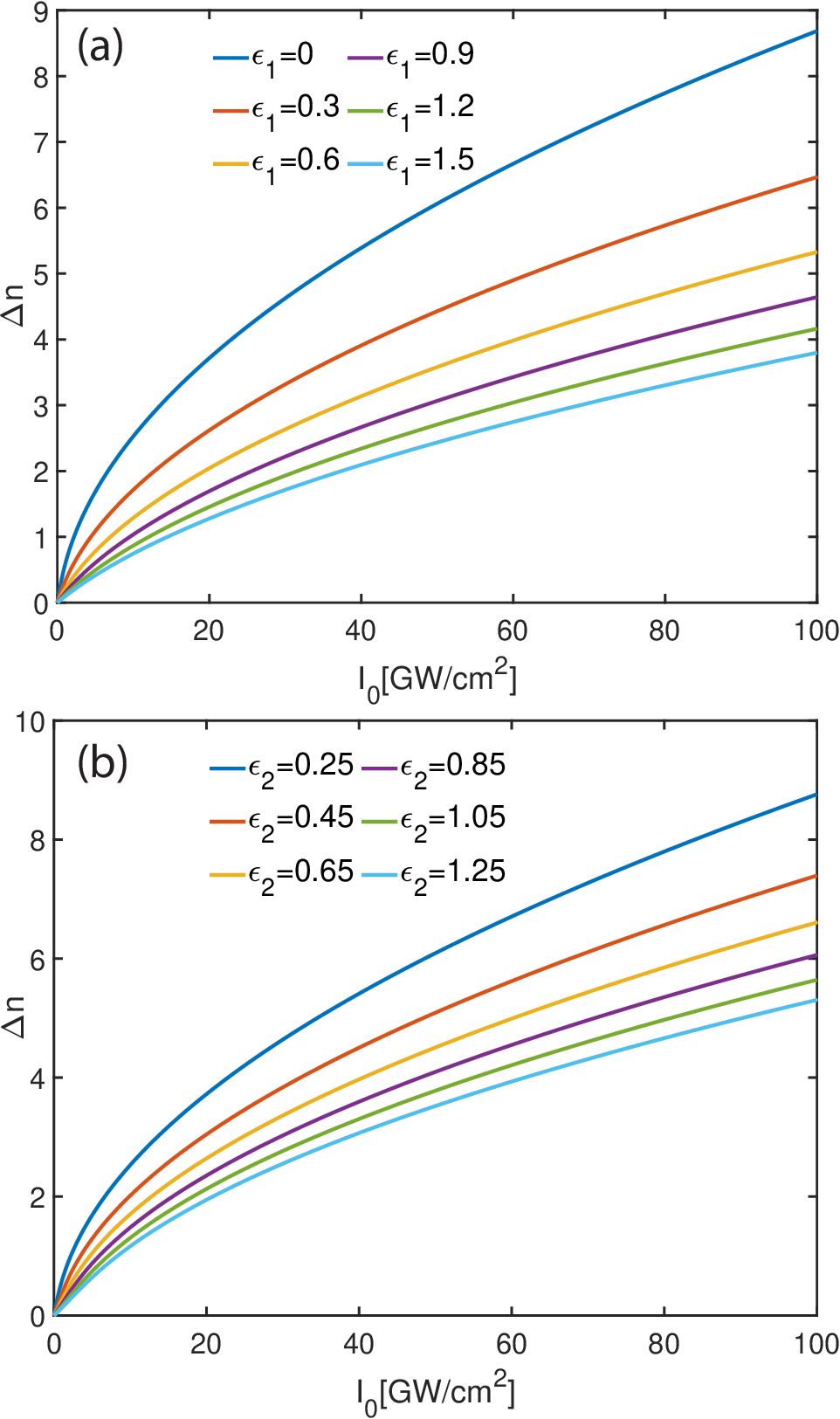}
\caption{\label{fig:deltan} Nonlinear refractive index change as a function of the peak pump intensity. (a) Effect of varying the real part of the linear permittivity away from the ENZ condition (i.e., $\epsilon_{1}=0$) for a fixed value of the imaginary part $\epsilon_{2}=0.25$ and (b) effect of varying the imaginary part of the linear permittivity (i.e., $\epsilon_{2}$) when $\epsilon_{1}=0$. In all cases we considered a real third-order susceptibility $\chi^{(3)}=5.2 \times 10^{-17}$ m$^{2}$/V$^{2}$ as measured in Ref. \cite{Tornike23}}
\end{center}
\end{figure}
It has been recently established by Reshef et al. \cite{Reshef2017}  that for ENZ low-index media the conventional expression  for the intensity-dependent refractive index in Eq. (\ref{nlindex}) is inapplicable as the nonlinear field dependence of the refractive index cannot be expressed in a perturbation series. Instead, the full non-perturbative expression below must be considered in the study of ENZ nonlinear media \cite{Reshef2017}:
\begin{equation}\label{nonpert}
	{\tilde{n}(\omega)}=\sqrt{\epsilon_{fc}(\omega)}=\sqrt{\epsilon_{r}^{(1)}+\sum_{j({odd})\neq{1}}c_{j}\chi^{(j)}|E(\omega)|^{j-1}},
\end{equation} 
where we introduced the complex nonlinear refractive index $\tilde{n}(\omega)\equiv{n+i\kappa}$ to emphasize that the quantities in Eq. (\ref{nonpert}) are generally complex and $c_{j}$ are the degeneracy for each relevant nonlinear process considered in the summation. Specifically, for ITO materials excited at large pump intensity up to $\approx$ 250 GW/cm$^{2}$, it was shown that nonlinear susceptibility up to seventh-order should be included in Eq. (\ref{nonpert}) because their contributions far exceed the linear refraction \cite{Reshef2017}.
\begin{figure}
\begin{center}
\includegraphics[width=0.4 \textwidth]{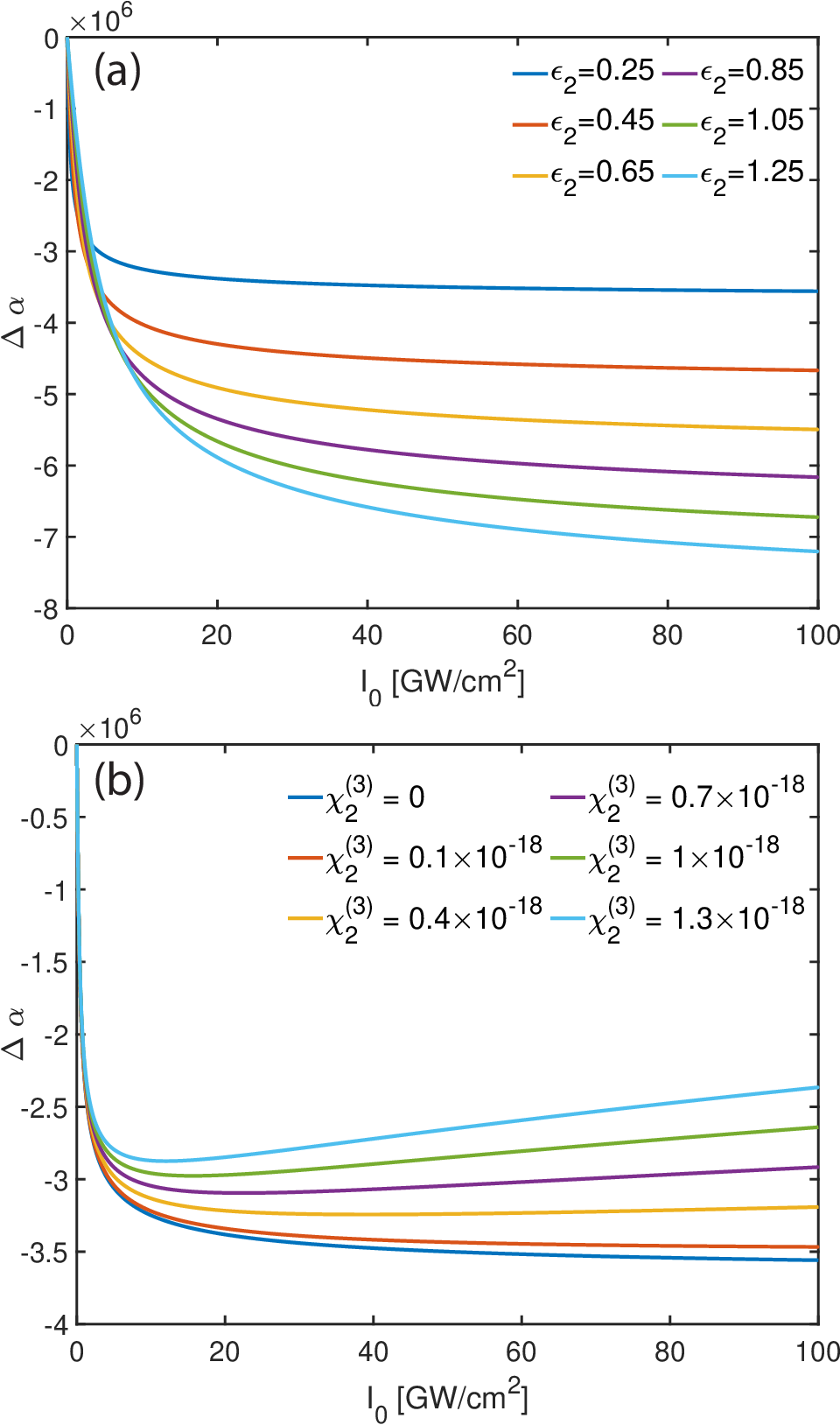}
\caption{\label{fig:deltalpha} Nonlinear absorption change as a function of the peak pump intensity. (a) Effect of varying the imaginary part of the linear permittivity (i.e., $\epsilon_{2}$) when $\epsilon_{1}=0$. (b) We considered a fixed $\epsilon_{2}=0.25$ and varied the imaginary part of the complex third-order susceptibility $\chi_{2}^{(3)}$. In all cases the real part of the third-order susceptibility was fixed to $\chi^{(3)}=5.2 \times 10^{-17}$ m$^{2}$/V$^{2}$ as measured in Ref. \cite{Tornike23}}
\label{figure2}
\end{center}
\end{figure}

\noindent Closed-form expressions for the real ($n$) and imaginary ($\kappa$) parts of the complex refractive index (and hence of the nonlinear absorption coefficient $\alpha=2k_{0}\kappa$, where $k_{0}=2\pi/\lambda$ is the free space wave number) can be obtained by inserting the appropriate perturbative terms of the field-corrected dielectric function inside the generally valid relations linking the real and imaginary parts of the complex refractive index, as shown in Ref. \cite{Tornike23}. In particular, if we consider low excitation intensities up to $\approx$ 10 GW/cm$^{2}$, where only third-order susceptibility terms are important, we obtain the following expressions:
\begin{subequations}\label{BH2}
\begin{eqnarray}
n &=& \sqrt{\frac{|\epsilon|+\epsilon^{(1)}_{1}+3\chi^{(3)}_{1}|E|^{2}}{2}},\label{BH2_n}\\
\alpha &=& \frac{4\pi}{\lambda}\sqrt{\frac{|\epsilon|-\epsilon^{(1)}_{1}-3\chi^{(3)}_{1}|E|^{2}}{2}},\label{BH2_kappa}
\end{eqnarray}
\end{subequations}
where: 
 \begin{equation}
 |\epsilon| = \sqrt{(\epsilon^{(1)}_{1}+3\chi^{(3)}_{1}|E(\omega)|^{2})^{2}+(\epsilon^{(1)}_{2}+3\chi^{(3)}_{2}|E(\omega)|^{2})^{2}}.
 \end{equation}
Here  $(\chi_{1},\chi_{2})$ denote, respectively, the real and imaginary parts of the susceptibility function \cite{boyd2008nonlinear}.
This approach can be naturally generalized to any desired order in the perturbation expansion of the susceptibility, which is important when studying ENZ materials at large pump intensities. 

\noindent In order to clearly illustrate the importance of ENZ low-index materials for nonlinear optics applications, we show in Figures \ref{fig:deltan} (a-b) the computed nonlinear refractive index change as a function of the peak pump intensity for different values of the real (a) and imaginary (b) parts of the linear permittivity of the medium. We considered susceptibility values up to $\chi^{(7)}$ that are typical of high-quality ITO films at their ENZ transition wavelengths, as measured in Refs. \cite{Tornike23,Reshef2017}. We see that the largest refractive index changes are always obtained when $\epsilon_{1}=0$ and the linear losses ($\epsilon_{2}$) are minimized. We also considered in Figures \ref{fig:deltalpha}  the variation of the nonlinear absorption coefficient as a function of peak intensity for different values of the linear losses (a) and of nonlinear losses (b) associated to $\chi_{2}^{(3)}$ at the ENZ condition defined by $\epsilon_{1}=0$. These saturation trends of the nonlinear losses have been recently observed experimentally in highly nonlinear ITO nanolayers excited around the $\lambda_{ENZ}$ wavelength  \cite{Tornike23,Reshef2017}.

Before proceeding further with the main result of our work, it is worth pointing out that while Eq. \eqref{nonpert} is a valid model for describing nonperturbative effects in low-index media, and that Eqs. \eqref{BH2_n} and \eqref{BH2_kappa} are only valid in the low-intensity regime. Therefore, extrapolating information on the high-intensity regime from those equations can lead to erroneous predictions. This is due to the fact that Eqs. \eqref{BH2_n} and \eqref{BH2_kappa} only includes the third-order susceptibility and completely neglect the higher-order terms in Eq. \eqref{nonpert}. These terms, as it is pointed out in Ref. \cite{Reshef2017}, however, play a significant role when the peak intensity of the pump pulse becomes very large. We will discuss in detail how to overcome this problem in Sect. \ref{saturated}.

\section{The Dressed Electromagnetic Field Lagrangian}\label{theory1}
\noindent In this section, we introduce a Lagrangian approach to light-matter interaction, originally proposed by Huttner and Barnett in 1992 \cite{huttner}, then extended to the case of dispersive media by Bechler in 1999 \cite{bechler}, and recently generalised to quantum nonlinear optics \cite{PhysRevA.100.053845}. The approach we develop in this work follows these references, and consists of the following steps: (i) we define the model through the Huttner-Barnett-Bechler Lagrangian; (ii) we calculate the dressed (effective) electromagnetic action via Feynman path integrals; (iii) we derive the expression for the displacement vector in frequency domain, and finally (vi) we extract the information on the dielectric function. 
A summary of the details for the calculation of the effective Lagrangian is given in Appendix \ref{appendixB}, and we refer the interested reader to Refs. \cite{bechler,PhysRevA.100.053845} for a more complete and detailed discussion of the model.
\begin{figure}
\begin{center}
\includegraphics[width=0.45 \textwidth]{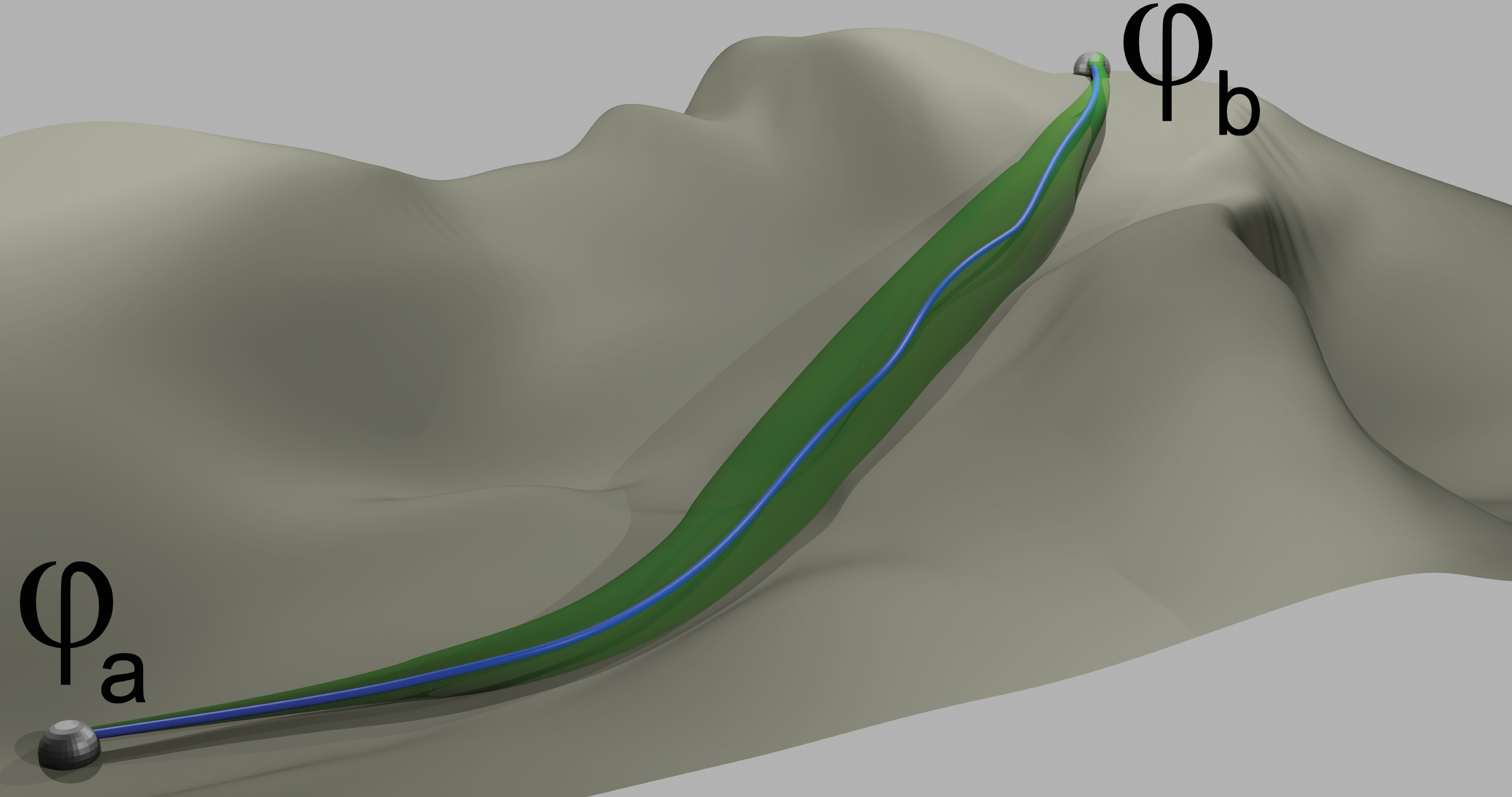}
\caption{Pictorial representation of path integrals for fields. The final field configuration $\varphi_b$ can be reached from the initial one ($\varphi_a$) by summing over all the possible field configurations (paths in field space, pictorially represented by the landscape). Each path is assigned a weight function $\exp\left[i\,S/\hbar\right]$, where $S$ is the classical action of the field. The classical equations of motions of the field, represented here by a solid, blue line, are those minimising the action.  The dark green region around it represents the region where the interference of all the contributing paths linking the initial ($\varphi_a$) and final ($\varphi_b$) field configurations is maximum.}
\label{figure3}
\end{center}
\end{figure}
To begin with, let us consider the action: $S=\int\,dt\,d^3\vett{x}\,\mathcal{L}_{HBB}$, with $\mathcal{L}_{HBB}=\mathcal{L}_{em}+\mathcal{L}_{matter}+\mathcal{L}_{res}+\mathcal{L}_{lmi}$ being the Huttner-Barnett-Bechler (HBB) Lagrangian \cite{huttner,bechler},
where $\mathcal{L}_{em}=\varepsilon_0|\vett{E}(\vett{x},t)|^2/2-|\vett{B}(\vett{x},t)|^2/(2\mu_0)$ is the standard Maxwell Lagrangian \cite{jackson} for the free electromagnetic field. Here, $\mathcal{L}_{matter}$ accounts for the dynamics of the matter (polarization) field $\vett{P}(\vett{x},t)$, modelled as an harmonic oscillator field with a characteristic resonance frequency $\omega_0$ and a static polarizability $\beta$, and $\mathcal{L}_{res}$ refers instead to the reservoir field, representing the collection of all decay processes occurring in the material and modelled as a collection of harmonic oscillators interacting with the matter field. The explicit expressions of the matter and reservoir Lagrangians are given in Appendix \ref{appendixA0}. The last term, $\mathcal{L}_{lmi}$ is the light-matter interaction Lagrangian, describing the interaction between the electromagnetic field and the matter field. Without loss of generality, we consider the interaction in the usual electric dipole approximation, i.e., $\mathcal{L}_{lmi}=-\vett{E}(\vett{x},t)\cdot\vett{P}(\vett{x},t)$.

The dressed electromagnetic Lagrangian, containing all the information on the optical properties of the medium, can then be calculated from the action above by eliminating the matter degrees of freedom through Feynman path integration \cite{feynmanHibbs, APLtutorial} as follows:
\beq\label{eq5t}
\int\,\mathcal{D}P\,\mathcal{D}R\,\exp\left[\frac{i\,S}{\hbar}\right]\equiv\,\exp\left[\frac{i\,S_{eff}}{\hbar}\right],
\eeq
where the integration measure $\mathcal{D}\varphi$ indicates path integration over all possible configurations of the field $\varphi$ \cite{srednicki,feynmanHibbs}. A pictorial representation of path integrals and their physical meaning is presented in Fig. \ref{figure3}. The procedure to calculate the effective dressed electromagnetic Lagrangian is sketched in Appendix \ref{appendixB} and its final expression, suppressing the $\vett{x}$-dependence for convenience, is given as follows:
\beq\label{eq16t}
\mathcal{L}_{eff}=\mathcal{L}_{em}+\int\,d\tau\,\vett{E}^*(t)\cdot\boldsymbol{\Gamma}(t-\tau)\cdot\vett{E}(\tau),
\eeq
where $\boldsymbol\Gamma(t-\tau)$ is the (tensorial) Green's function of the material \cite{bechler}, i.e., the optical response of the material. 
\subsection{Effective Lagrangian in Fourier Space}
\noindent To derive the dielectric function of the medium, it is useful to represent the effective action in frequency domain by using Fourier transformation to obtain $S_{eff}=\int\,(d\omega/2\pi)\,dx\,\tilde{\mathcal{L}}_{eff}$, where now $\tilde{\mathcal{L}}_{eff}=\varepsilon_0|\vett{E}(\omega)|^2+|\vett{B}(t)|^2/(2\mu_0)+\vett{E}^*(\omega)\cdot\tilde{\Gamma}(\omega)\cdot\vett{E}(\omega)$, 
and $\tilde{\Gamma}(\omega)$ is the Fourier transform of $\boldsymbol\Gamma(t-\tau)$, whose explicit expression is \cite{bechler}:
\beq\label{eq18t}
\tilde{\Gamma}(\omega)=\frac{\varepsilon_0\omega_0^2\beta}{\omega_0^2-\omega^2\left[1+\lambda(\omega)\right]}.
\eeq
Here, $\lambda(\omega)$ is a spectral function that contains information about the properties of the medium and the spectral coupling to the reservoir \cite{bechler}, i.e., the linear losses of the material. Choosing an explicit expression for $\lambda(\omega)$ conveniently allows for different models of dispersive materials.
From the effective Lagrangian, we can then derive the electric displacement vector as \cite{landauCM} $\vett{D}(\omega)=\partial\tilde{\mathcal{L}}_{eff}/\partial\,\vett{E}^*(\omega)\equiv\,\varepsilon_0\,\varepsilon(\omega)\,\vett{E}(\omega)$, from which we define the linear dielectric function as $\varepsilon(\omega)=1+\tilde{\Gamma}(\omega)/\varepsilon_0\equiv 1+\chi(\omega)$.
This result provides the possibility for the model to describe different classes of materials by suitably choosing different values of $\lambda(\omega)$.
\subsection{ENZ Materials}\label{sect43}
\noindent The dielectric permittivity of an ENZ material around its $\lambda_{ENZ}$ is derived in a similar manner to that of a Drude metal, with the inclusion of losses and a background permittivity $\varepsilon_{\infty}$, due to the residual matter polarization of the ion cores serving as background for the dynamics of the nearly free electrons of the material \cite{maier_plasmonics_2007}. This leads to the following expression for the dielectric function of an ENZ material around the $\lambda_{ENZ}$:
\beq \label{DrudeLorentz}
\varepsilon(\omega)=\varepsilon_{\infty}-\frac{\omega_p^2}{\omega(\omega\pm i\gamma)},
\eeq
where $\gamma$ accounts for the losses, and the sign at the denominator depends on the adopted convention. The second term of the dielectric function above can be readily obtained from our model, by setting $\lambda(\omega)=(\omega_0^2\pm i\gamma\omega)/\omega^2$. To account for the first term, instead, we need to consider a slightly modified form of the Lagrangian that includes the residual polarization $\vett{P}_{\infty}(\vett{x},t)=[\varepsilon_0(\varepsilon_{\infty}-1)/2]\vett{E}(\vett{x},t)$ due to the ion cores. To do so, we introduce this term in the light-matter interaction Lagrangian 
$\mathcal{L}_{lmi}=-\vett{E}(\vett{x},t)\left[\vett{P}(\vett{x},t)+\vett{P}_{\infty}(\vett{x},t)\right]$. Since the term $\vett{P}_{\infty}(\vett{x},t)$ is constant with respect to the matter field $\vett{P}(\vett{x},t)$, we can immediately bring it out of the path integrals and insert it directly in the expression for the effective Lagrangian. Including this modification into the effective Lagrangian, we get the correct form of the dielectric function for an ENZ material around its ENZ transition, i.e., Eq. \eqref{DrudeLorentz}.
\section{Nonlinear dielectric function: Perturbative Case}\label{theory3}
\noindent Nonlinear electromagnetic effects can be introduced in the model by choosing an interaction Lagrangian of the form $\mathcal{L}_{int}=\sum_{n}c_n\,(|\vett{E}(t)|^2)^{n}$, where the nonlinear couplings $c_n$ can be real or complex, and their explicit expression is determined by the particular kind of nonlinearity. Normally, $|c_n|\ll 1$ and the interaction is perturbative. This, for example, is the case of standard nonlinear optics \cite{boyd2008nonlinear}. Adding this interaction term into the HBB Lagrangian allows us to write the nonlinear (perturbative) dielectric function as:
\beq\label{eq19tNew}
\varepsilon(\omega,E)=1+\frac{1}{\varepsilon_0}\tilde{\Gamma}(\omega)+\frac{1}{\varepsilon_0\,\vett{E}(\omega)}\frac{\partial\tilde{\mathcal{L}}_{int}}{\partial\,\vett{E}^*(\omega)}.
\eeq
\subsection{Nonlinear Refraction}\label{SecPert}
\noindent As an example to familiarise with this formalism, let us derive the standard intensity law for standard Kerr materials, i.e., $n=n_0+n_2I$\, \cite{boyd2008nonlinear}. The interaction Lagrangian for a Kerr medium is \cite{srednicki,drummond}:
\beq
\mathcal{L}_{int}=\frac{3}{2}\varepsilon_0\,\chi^{(3)}\left|\vett{E}(t)\right|^4
\eeq
where $\chi^{(3)}$ is the (scalar) third-order nonlinear susceptibility. To make calculations easier, from this point on we consider a scalar electromagnetic field (i.e., we neglect phase matching and the effect of polarisation), and assume that the electric field only depends on one spatial coordinate, i.e., $\vett{E}(\vett{x},t)\rightarrow\,E(x,t)$. Then, for convenience, we suppress the spatial dependence, since it does not play any role in this calculation. Following the prescriptions of Sect. \ref{theory1}, we first represent the interaction Lagrangian in frequency domain (see Appendix \ref{appendixC}), then, using Eq. \eqref{eq19tNew} we get the nonlinear dielectric function as:
\beq\label{eq36t}
\varepsilon(\omega,E)=\varepsilon_{lin}(\omega)+3\chi^{(3)}\int\,d\Omega\,|E(\Omega)|^2,
\eeq
If we now introduce the time-averaged field intensity as \cite{boyd2008nonlinear} $I=2c\,n\,\varepsilon_0\,\,|E(\Omega)|^2$  and define the refractive index as $n=\Re\{\sqrt{\varepsilon(\omega,E)}\}$, we get, using the fact that $|\chi^{(3)}|\ll 1$, the well-know result \cite{boyd2008nonlinear}  $n\simeq\,n_0+n_2\,I+\mathcal{O}(I^2)$,
where $n_0=\Re\{\sqrt{\varepsilon_{lin}}\}$ is the linear refractive index, and $n_2$ is the usual nonlinear refractive index \cite{boyd2008nonlinear}, in accordance with the results obtained in Sect. \ref{ENZintro}
\section{Born-Infeld Lagrangian and Non-perturbative Nonlinearities}\label{section5}
\noindent In this section we propose a new approach to describe nonlinear materials with fully non-perturbative optical responses. The approach developed in Sect. \ref{ENZintro} already provides a partial solution to this problem, by considering the full non-perturbative action of the square root in the definition of the refractive index (see, for example, Eq. \eqref{nonpert}). However, the calculation of the dielectric function is still carried out with a standard, perturbative approach. Going beyond this requires to rethink the way nonlinearities are treated at the field theory level. The approach used above, for example, fails to describe a non-perturbative nonlinearity, since every nonlinear interaction is inserted ad-hoc in a perturbative manner, and even if one would take into account all the infinite orders of nonlinearities, the interaction Lagrangian cannot be easily resummed into a closed-form, non-perturbative expression. 

For these reasons, a paradigm shift in the way we think about optical nonlinearities is necessary. What we are looking for is a generalized electromagnetic theory, which still has Maxwell's equations as its dynamical equations of motion, but that at the same time can automatically account for optical nonlinearities.
Remarkably, such a theory was originally proposed in 1934 by Born and Infeld that developed a classical electrodynamics approach to solve the issue of the diverging electron self-energy \cite{bornInfeld,Sommerfeld}. The Lagrangian they proposed to describe such theory is known nowadays as the Born-Infeld (BI) Lagrangian and has the following expression \cite{Alam_2022,Sorokin_2022}:
\beq\label{lagBI}
\mathcal{L}_{BI}=b^2\left(1-\sqrt{1-\frac{\varepsilon_0|\vett{E}|^2-\frac{1}{\mu_0}|\vett{B}|^2}{b^2}}\right),
\eeq
where $b$ is a positive real constant, with the dimensions of an energy density, whose physical meaning is that of an upper limit to the energy carried by the field and, ultimately, its amplitude. The BI Lagrangian is the archetype of nonlinear electrodynamics in classical field theory (see, for example, the discussions in Refs. \cite{Sorokin_2022,Alam_2022}), and has recently experienced a resurgence in various areas of theoretical physics, including cosmology \cite{cosmology}, nuclear physics \cite{nuclearBI}, and string theory, where the BI Lagrangian governs the electrodynamics on D-branes \cite{zweibach}. 

A peculiar property of the BI Lagrangian is that its equations of motion are still Maxwell's equations, where now the electric and magnetic fields are intrinsically nonlinear. For this reason, the usual constitutive relations $\vett{D}=\varepsilon_0\vett{E}+\vett{P}$ and $\vett{H}=\vett{B}/\mu_0-\vett{M}$ need to be replaced with their nonlinear counterparts \cite{Sorokin_2022, birula}
\begin{subequations}
\begin{align}
\vett{D}&=\left(\frac{1}{\sqrt{1-\frac{\varepsilon_0|\vett{E}|^2-\frac{1}{\mu_0}|\vett{B}|^2}{b^2}}}\right)\,\vett{E},\\
\vett{H}&=\left(\sqrt{1-\frac{\varepsilon_0|\vett{E}|^2-\frac{1}{\mu_0}|\vett{B}|^2}{b^2}}\right)\,\vett{B}.
\end{align}
\end{subequations}
Moreover, in the so-called strong limit $b\rightarrow\infty$, the BI Lagrangian converges to the usual Maxwell Lagrangian, and the nonlinear interactions naturally arise as higher order terms in the expansion with respect to $b$.
\begin{figure}[!t]
\begin{center}
\includegraphics[width=0.4 \textwidth]{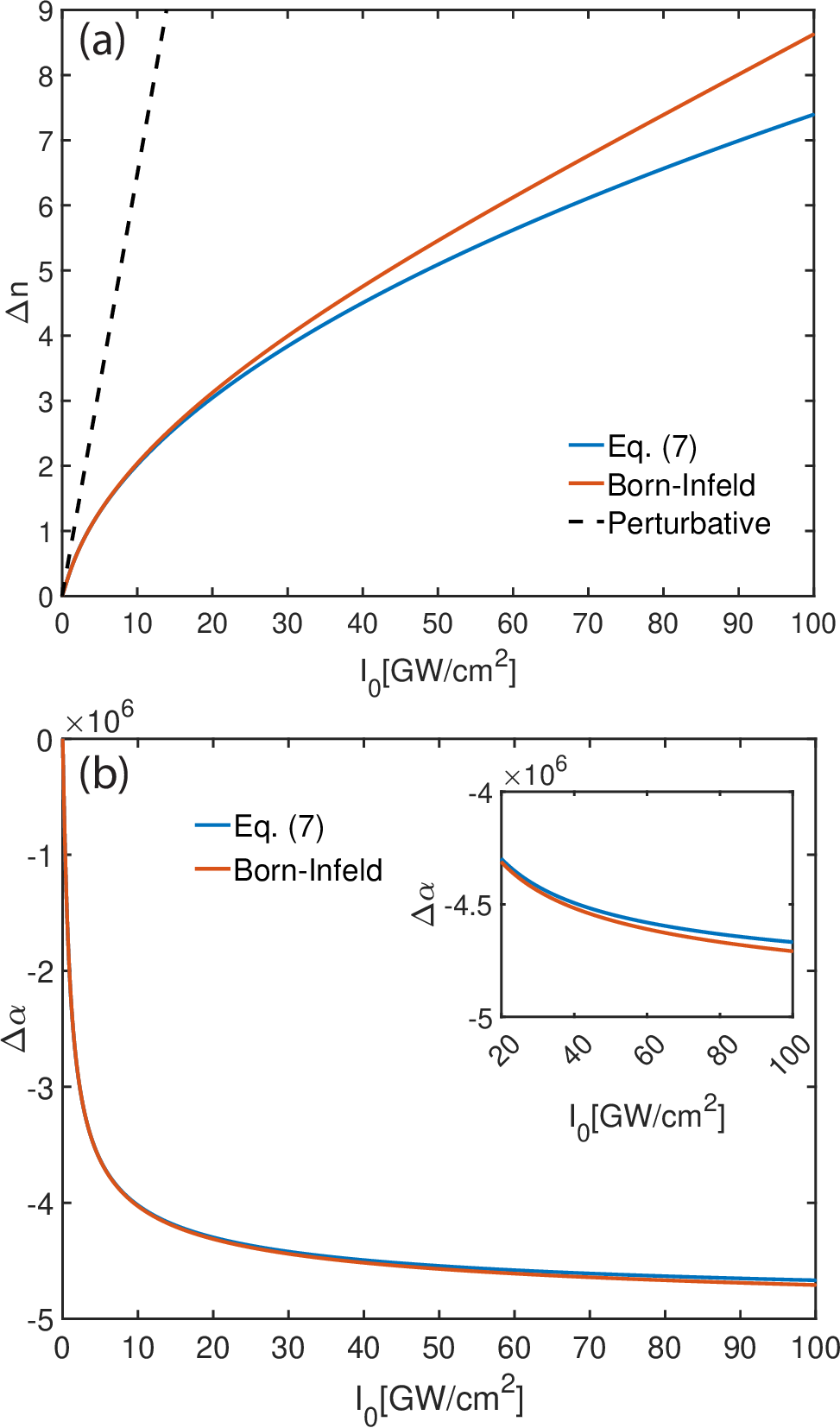}
\caption{(a) Nonlinear refractive index change and (b) nonlinear absorption as a function of the peak intensity as calculated using the Born-Infeld model (blue, solid line) and the non-perturbative refractive index  model in Eq. \eqref{nonpert} (red, solid line). As it can be seen, although the two models agree at low intensities, for $I_0\gtrsim 20$ $\text{GW/cm}^2$, the two models start to deviate significantly. The black, dashed line in panel (a), representing the perturbative approximation $\Delta\,n=n_2\,I_0$, has been inserted for comparison. To plot these curves, the following experimentally measured parameters (see Ref. \cite{Tornike23}) have been used: $\varepsilon_1=0$, $\varepsilon_2=0.45$, and $\chi^{(3)}=7.58\times 10^{-17}$ $\text{m}^2/\text{V}^2$. With these parameters we get $n_0=0.4743$ and $n_2=0.9517$ $\text{cm}^2/\text{GW}$. The BI model has been plotted using $I_c=300$ $\text{GW/cm}^2$. }
\label{figure4}
\end{center}
\end{figure}
\subsection{The Non-perturbative Dielectric Function }\label{epsNP}
\noindent Motivated by these arguments, we then propose to use a BI-like model to describe the non-perturbative optical nonlinearities of materials. In order to achieve this goal, we replace the Maxwell Lagrangian $\mathcal{L}_{em}$ and the interaction Lagrangian $\mathcal{L}_{int}$ in the nonlinear HBB model with the following Lagrangian:
\beq
\mathcal{L}=a\,\mathcal{L}_{BI}-\frac{a\,\varepsilon_0}{2}|\vett{E}(t)|^2,
\eeq
where $a$ is a parameter that we introduce to account for the fact that we are describing the dynamics of the electromagnetic field inside a medium and not in vacuum. The second term serves to ensure that the linear dielectric function can be written as $\varepsilon_{\infty}+\tilde{\Gamma}(\omega)/\varepsilon_0$. Notice, moreover, that since in nonlinear optics we are typically only interested on the nonlinearities caused by the electric field, the magnetic part of the BI Lagrangian plays no role in this setting, and can be ignored. 
By doing so, the effective non-perturbative Lagrangian now reads:
\begin{eqnarray}\label{eq28}
\mathcal{L}_{eff}^{(np)}=a\,\mathcal{L}_{BI}+\frac{\varepsilon_0(\varepsilon_{\infty}-a)}{2}|E(t)|^2+ \\ \nonumber 
+ \int\,d\tau\,E^*(t)\Gamma(t-\tau)E(\tau).
\end{eqnarray}

\noindent To calculate the non-perturbative dielectric function, we need to switch to a frequency-domain representation. However, the presence of the square root term in the BI Lagrangian makes this step particularly complicated, as one would require to find a suitable Fourier representation, which is generally not an easy task. One possible way to tackle this problem is to write the BI Lagrangian as a power series in $b$, represent each term of the series in frequency domain, and then re-sum the series. If we do so, and follow the procedure highlighted in Appendix \ref{appendixD}, we get the following result for the nonlinear dielectric function:
\beq\label{npEps}
\varepsilon^{(np)}(I)=\varepsilon_{lin}+\,\frac{a}{2}\,\left(\frac{I/Ic}{\sqrt{1-I/I_c}}\right)
\eeq
where $\varepsilon_{lin}$ is the linear permittivity of the material, and $I_c=2cn_0b^2$ is the critical intensity. The physical meaning of the critical intensity follows from the parameter $b$ originally introduced by Born and Infeld to set a limit on the maximum electric field when the magnetic flux $\textbf{B}=0$. This feature, which is a general characteristic of the BI electrodynamics, sets a maximum intensity scale to the static electric field in order to avoid the divergence of the electron self-energy (see, for example, Refs. \cite{bornInfeld,Sorokin_2022,Alam_2022} for a more detailed discussion of this point). In our context, however, one could interpret $I_c$ as an intrinsic free parameter of the BI model that can be used to fit the experimental data of the particular material considered. A caveat, however, is necessary: $I_c$ cannot be chosen completely arbitrarily, but it must be chosen in such a way that $1-I/I_c>0$ is fulfilled in order to avoid nonphysical solutions potentially arising when the square root denominator in Eq. (\ref{npEps}) becomes imaginary.

This is the main result of this work. Using the BI Lagrangian enables a novel, non-perturbative approach to describe the optical response of strongly nonlinear media. The price to pay, however, is that intrinsically perturbative parameters such as the nonlinear susceptibilities loose their meaning in this broader context, but can be conveniently restored in the perturbative limit of the BI theory by setting $b\rightarrow\infty$.
\subsection{Nonlinear Response of ENZ Low-index ITO}
\noindent We now compare the two models described in this work, i.e., the non-perturbative BI model above, and the one derived in Sect. \ref{ENZintro}. To this aim, we use ITO as the nonlinear ENZ material, for which the parameters appearing in Eq. \eqref{DrudeLorentz} have been extracted from fitting the dielectric function model of ITO described in Ref. \cite{Tornike23} around $\lambda_{ENZ}$ and are $\varepsilon_{\infty}=3.99$, $\omega_p=3.16\times 10^{15}$ Hz, and $\gamma=0.185\times 10^{15}$ Hz. We take as a figure of merit for the comparison the nonlinear refractive index change $\Delta\,n=n(I)-n(0)$, and the nonlinear absorption change $\Delta\alpha=\alpha(I)=\alpha(0)$.
We assume an impinging electromagnetic field oscillating at frequency $\omega=1.57\times 10^{15}$ Hz, corresponding to the ENZ wavelength of ITO, i.e., $\lambda=1200$ nm. 

Figure \ref{figure4} shows the comparison between the non-perturbative approach developed in Sect. \ref{ENZintro} (blue, solid curve), with the BI model of Sect. \ref{epsNP} (red, solid curve). To match the latter to the former, we require that they agree in the small intensity regime, where the BI model predicts $\varepsilon^{(np)}\simeq\,\varepsilon_{lin}+a\,I/(2I_c)+\mathcal{O}(I^2)$ and the usual permittivity model for Kerr nonlinearities is $\varepsilon\simeq\varepsilon_{lin}+3\chi^{(3)}|E|^2$. This leads to identifying $a=4n_0n_2I_c$. With this fixed, $I_c$ remains the only free parameter of the BI model. 
\begin{figure}[!t]
\begin{center}
\includegraphics[width=0.5 \textwidth]{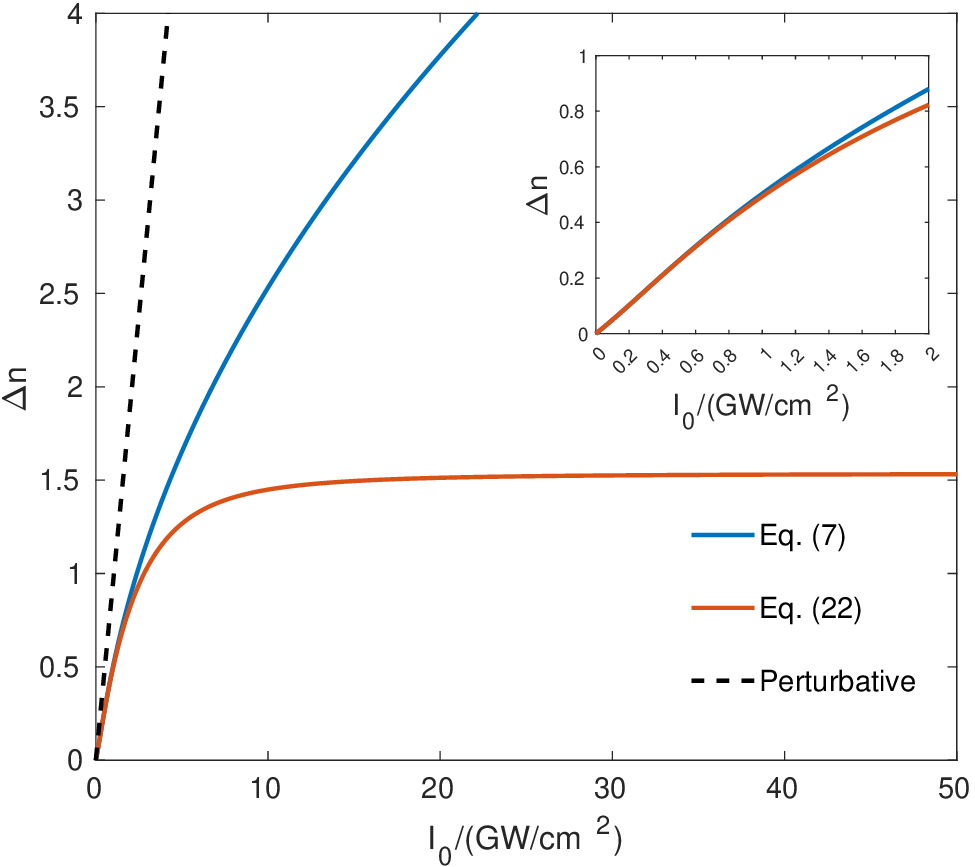}
\caption{Nonlinear refractive index change as a function of the peak intensity calculated using the saturated (red, solid line) Born Infeld model, compared with the nonperturbative refractive-index model in Eq. \eqref{nonpert} (blue,solid line). Both models agree at low intensities (see inset), but while the non-perturbative model predicts an ever increasing refractive index change, the saturated model of Eq. \eqref{npEpsSat} correctly reproduces the physical behaviour at high itensities. The parameters used to plot these curves are the same as in Fig. \ref{figure4}. For the saturated permittivity, we have taken $\varepsilon_{sat}=4$, compatible with the experimental measurement of ITO presented in Ref. \cite{Tornike23} the black, dashed line, representing the perturbative approximation $\Delta n=n_2 I$, has been inserted for comparison.}
\label{figure5}
\end{center}
\end{figure}
Figure \ref{figure4}(b) shows the nonlinear absorption calculated from the BI model (red, solid line) compared with the results from Eq. \eqref{nonpert} (blue, solid line) for the case $\chi_2^{(3)}=0$, i.e., assuming that the imaginary part of the third-order susceptibility is zero. The condition $\chi_2^{(3)}=0$ is a necessary one for the BI model. Contrary to standard nonlinear optics, where the susceptibility is in general a complex quantity, the free parameter of the BI model, namely the critical intensity $I_c$, is instead an intrinsically real parameter. This limits the possibility for the BI, in its actual form, to correctly reproduce the behaviour of Fig. \ref{figure2}(b), when $\chi_2^{(3)}\neq 0$. As it is discussed below, however, including this feature in the BI model might be possible by introducing extra terms in the Lagrangian, giving rise to an effective field theory with a complex free parameter.
\subsection{Nonlinear Saturation Effects}\label{saturated}
\noindent The model proposed in Eq. \eqref{nonpert} introduces the susceptibility as a scaling factor between the polarization and the incident field. Although in principle this effective macroscopic approach is legitimate and the effective susceptibility defined with this model can be used to readily compare the effective and microscopic models, one should be careful in doing so. In fact, for this comparison to be valid, one should ensure that the effective description remains within the bounds of the microscopic process and the experimental conditions upon which such a model is based. In fact, using the effective $\chi^{(3)}$ model incorrectly, could lead to erroneous results.

In the case of ENZ materials, for example, it is generally accepted that the intensity-dependent refractive index arises from absorptive effects and free electrons \cite{tice2014ultrafast, kinsey2015epsilon, Alam2016, ramesh2019creating, un2023electronic} This implies the existence of a finite relaxation time for the nonlinear susceptibility or, equivalently, the intensity-dependent refractive index, which makes the values of the nonlinearity significantly dependent upon the duration of the excitation pulse \cite{reichert2014temporal}. As a result, the nonlinear response of ENZ materials will eventually saturate to a value corresponding to the optical response of the material in the absence of the free-electron Drude dynamics. This saturation process has been discussed and experimentally verified under different circumstances and for different ENZ materials \cite{Alam2016,benis2022extremely,caspani2016enhanced,wang2019extended, Britton2022}.

To correctly take into account this saturation effect starting from Eq. \eqref{nonpert}, one possible strategy is to introduce higher order susceptibilities, which progressively become more relevant as the intensity increases. This is the approach taken, for example, in Ref. \cite{Reshef2017}. To account for this mechanism in our intrinsically nonperturbative model, described by Eq. \eqref{npEps}, however, we cannot use this argument, since we do not have acces to the different orders of nonlinear susceptibility. Instead, we add a saturation term in our model, resulting in the following version of our non-perturbative model for the permittivity of ITO;
\beq\label{npEpsSat}
\varepsilon^{(np)}(I)=\varepsilon_{lin}+\,\frac{a}{2}\,\left(\frac{I/Ic}{\sqrt{1-I/I_c+(I/I_0)^2}}\right),
\eeq
where $I_0$ is the high-power saturation intensity, and its value can be determined by requiring that at high pump intensities the Drude component of the linear permittivity is completely saturated, and only the Lorentz-Tauc contribution remains to describe the ITO permittivity \cite{Tornike23, Britton2022}. Calling then $\delta\varepsilon=\varepsilon_{sat}-\varepsilon_{lin}$, where $\varepsilon_{sat}$ is the saturated permittivity at high intensity, the saturation intensity $I_0$ can be fixed by the behaviour of the above equation near infinity as $I_0=|\delta\varepsilon/2n_0n_2|$.

From a phenomenological point of view, the saturation of the nonlinearity of a material at high intensity arises from the fact that the nonlinear coupling constant itself acquires an intensity dependence at high values of $I$. Since in our nonperturbative model the critical intensity $I_c$ plays the role of the nonlinear coupling coefficient, one could insert this saturation mechanism by simply replacing a constant value of $I_c$ with an intensity-dependent critical intensity of the form $I_c(I)=I_c/[1-(I_c/I_0)I]$, so that for $I_0\rightarrow\infty$ one recovers the unsaturated case (i.e., low intensity), and for $I_0<I_c$ one can use the parameter $I_0$ to compensate the divergence in Eq. \eqref{npEps} and induce saturation.

From a more fundamental field theory perspective, on the other hand, the presence of the extra term in Eq. \eqref{npEpsSat} needs to be justified at the Lagrangian level by adding an extra term in the Born-infeld Lagrangian, leading to consider:
\beq
\mathcal{L}_{BI}^{(sat)}=b^2\left[1-\sqrt{1-\frac{\varepsilon_0^2|E(t)|^2}{b^2}+\frac{\alpha|E(t)^4|}{b^4}}\right],
\eeq
where the parameter $\alpha$ is related to $I_0$. In Born-Infeld nonlinear electrodynamics, a similar extra term is related to the dual electromagnetic tensor\cite{Alam_2022}, which is ultimately proportional to $(\vett{E}\cdot\vett{B})^2$. The presence of this extra term could also be justified by more complicated models of nonlinear electrodynamics, such as Dirca-Born-Infeld-like Lagrangians used to study gauge fields on D-branes \cite{zweibach}, which allow higher-order terms inside the square root in Eq. \eqref{lagBI} in terms of coupling of the electromagnetic field to background scalar fields.
\begin{figure}[!t]
\begin{center}
\includegraphics[width=0.5 \textwidth]{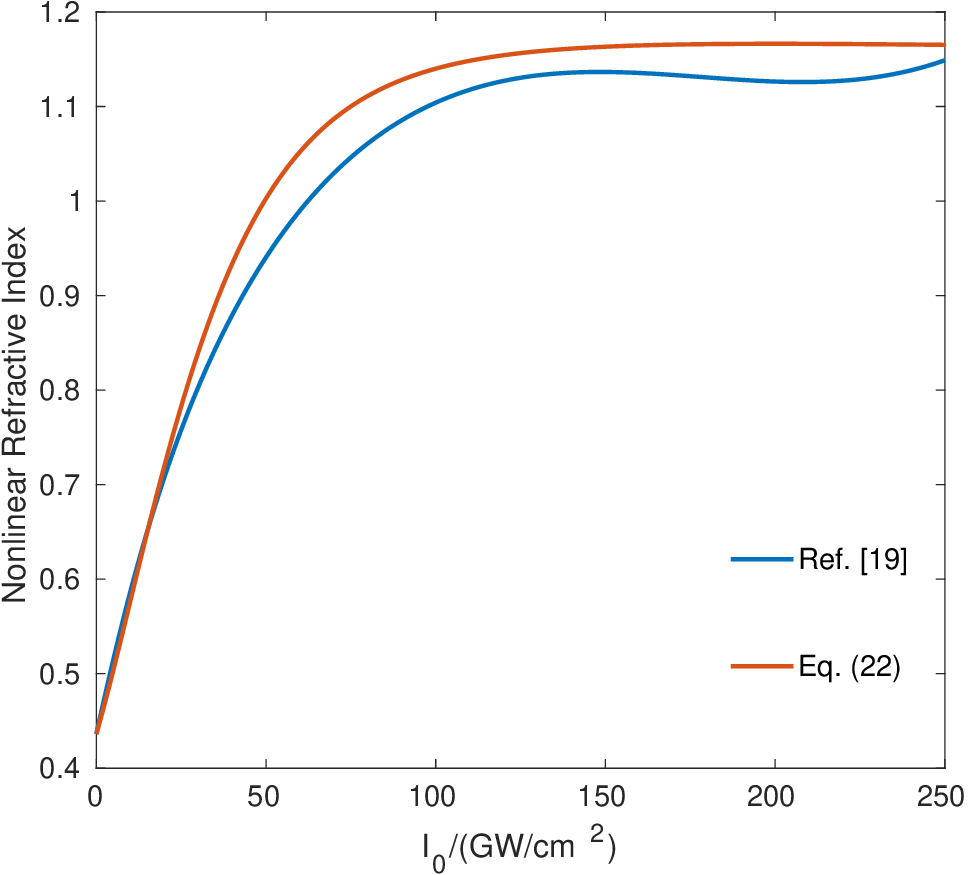}
\caption{Nonlinear refractive index as a function of the peak intensity caluclated using the saturated (red, solid line) Born-Infeld model, compared with the nonperturbative model with $5^{\text{th}}$ and $7^{\text{th}}$ order nonlinearities from Ref. \cite{Reshef2017} (blue, solid line). The parameters used to plot the saturated Born-Infeld model are $I_c=100$ GW/cm$^2$ and $\varepsilon_{sat}=1.252+i 0.606$, as taken from Ref.\cite{Reshef2017}. The values for the nonlinear susceptibilities for the blue curve have been taken from Table 1 in Ref. \cite{Reshef2017}.}
\label{figure6}
\end{center}
\end{figure}

In Fig. \ref{figure5} we compare the effective susceptibility model of Eq. \eqref{nonpert} and the saturated Born-Infeld model given by Eq.\eqref{npEpsSat}. As it can be seen, while the models agree very well at low intensities, where the nonlinear response is governed mainly by $I_c$ (see inset), at high intensity, the saturated model correctly predicts the saturation of the nonlinear index change, while the unsaturated one erroneously predicts ever increasing $\Delta\,n$ values. As an additional proof of the validity of our model, in Fig. \ref{figure6} we compare the predictions of Eq. \eqref{npEpsSat} with the non-perturbative, saturated refractive index model developed by Reshef and co-workers in Ref. \cite{Reshef2017}, where the summation under the square root symbol in Eq. \eqref{nonpert} includes also $5^{\text{th}}$ and $7^{\text{th}}$ order nonlinear susceptibility contributions. As it can be seen, our saturated BI model reproduces the experimental data from Ref. \cite{Reshef2017} quite well. Finally, it should be noted that the proposed saturation model does not take into account nonlocal and gradient terms in the nonlinear response of ITO as well as surface nonlinearities, and therefore cannot be directly applied to  the study of nonlinear nano-structures where larger refractive index modulations have been experimentally demonstrated \cite{Reshef2019_ENZ_Review,Tornike23,Alam2018}.


In this work we have only considered the impact of electric nonlinearities, neglecting the effect of the magnetic field. Taking this also into account might as well introduce new degrees of freedom, that can be used to tame the divergence. In recent years, several other models of nonlinear electrodynamics, such as NED or ModMax \cite{Sorokin_2022}, have been raising some interest in the field theory community. Adapting those models to non-perturbative nonlinear optics might also be interesting, as it could possibly lead to alternative perspectives. Lastly, it is worth noticing that when $b\rightarrow 0$, the BI Lagrangian becomes zero and the electromagnetic field in this limit has no dynamics \cite{birula3}. In this case, a Hamiltonian description of the electromagnetic field is necessary and shows the occurrence of topological objects, such as hopfions \cite{ranada} and knotted fields \cite{birula2}. Making the connection between the BI approach and nonlinear ENZ materials could therefore open new horizons for the study of electrodynamics in ENZ media, possibly hinting at the existence of hidden topological properties as well.

\section{Conclusion}\label{conclusions}
\noindent In this paper we have introduced a fully non-pertubative approach for the description of the nonlinear optical refraction properties of ENZ low-index media and non-perturbative metamaterials, which also includes the intrinsic nonlinear saturation effects originating from the light-induced bleaching of the Drude permittivity. In particular, we used the rigorous Feynman path integral method to develop an effective Lagrangian field theory for light-matter interactions and investigated, within a simple scalar model, the relevant case of Indium Tin Oxide (ITO) nonlinear materials. Remarkably, we established that their non-perturbative refractive index response naturally emerges from an
intrinsically nonlinear electrodynamics theory characterized by the Born-Infeld (BI) Lagrangian. At large optical intensities, the developed BI model naturally accounts for all orders of nonlinear susceptibilities of materials, unlike the case of traditional perturbative approaches. Moreover, we have shown that very concept of nonlinear susceptibility coefficients loses its meaning in the non-perturbative regime and must be substituted with the nonlinear dielectric function in Eq. \eqref{npEpsSat}, directly derived from the BI Lagrangian. A closed-form, two-parameter formula for the non-perturbative dielectric function was derived, containing all orders of nonlinearities of an arbitrary material. Our model goes beyond the state-of-the-art perturbative expansion of the refractive index, and introduces a new paradigm for calculating the dielectric function in a fully non-perturbative manner. These results extend the Huttner-Barnett-Bechler electrodynamics model to the case of non-perturbative photonic ENZ metamaterials and devices providing a nonlinear, field-theoretic framework for describing exceptional nonlinearities beyond traditional perturbation theory.

\begin{acknowledgments}
\noindent M.O. and Y.T. acknowledge the financial support from the Academy of Finland Flagship Programme (PREIN - decision Grant No. 320165) and the Academy of Finland project AQUA-PHOT (decision Grant No. 349350). L.D.N. and T.S. acknowledge the support of the U.S. Army Research Office under award number: W911NF2210110.
\end{acknowledgments}
\appendix
\section{Matter Lagrangians}\label{appendixA0}
The explicit expression of the matter Lagrangian $\mathcal{L}_{matter}$ used in the text is given as:
\beq
\mathcal{L}_{matter}=\frac{1}{2\varepsilon_0\omega_0^2\beta}\left[\dot{\vett{P}}(\vett{x},t)^2-\omega_0^2\vett{P}(\vett{x},t)^2\right],
\eeq
with $\omega_0$ being the characteristic resonance frequency of the material, $\beta$ its static polarizability, and the dot indicates derivative with respect to time. 

Analogously, the reservoir field is defined as a collection of harmonic oscillators, each with a characteristic frequency $\omega$, as follows:
\begin{eqnarray}
\label{eq3t}
\mathcal{L}_{res}&=&\int_0^{\infty}\,d\omega\,\Bigg\{\frac{1}{2}\left[\dot{\vett{R}}(\vett{x},t;\omega)^2-\omega^2\vett{R}(\vett{x},t;\omega)^2\right]- \\ \nonumber 
&-&f(\omega)\vett{P}(\vett{x},t)\cdot\dot{\vett{R}}(\vett{x},t;\omega)\Bigg\},
\end{eqnarray}
where $\vett{R}(\vett{x},t;\omega)$ is the reservoir field, and $f(\omega)$ is a spectral function depicting the frequency-dependence of the matter-reservoir coupling.
\section{Derivation of the Effective Lagrangian}\label{appendixB}
\noindent To derive the Effective Lagrangian in Eq. \eqref{eq16t}, we need to perform the two path integrations, with respect to the matter and reservoir fields, respectively. To this aim, it is convenient to first integrate with respect to the reservoir field $\vett{R}(\vett{x},t;\omega)$, since this integration does not contain any electric-field-dependent terms, and then with respect to the matter field $\vett{P}(\vett{x},t)$. The path integral procedure is the same in both cases, but slightly more complicated for the latter. Here, we present a detailed calculation of the first path integral, i.e., the one on $\vett{R}(\vett{x},t;\omega)$ and point the interested reader to Refs. \cite{bechler, PhysRevA.100.053845} for a complete derivation of the second path integral and more details on the procedure. For the sake of simplicity, we suppress the $\vett{x}$-dependence, and indicate it explicitly only when needed.

To start with, let us rewrite Eq. \eqref{eq16t} in the following way, isolating the two path integrals:   
\beq\label{eq8t}
\int\,\mathcal{D}P\,\mathcal{D}R\,\exp\left[\frac{i\,S}{\hbar}\right]=\exp\left[\frac{i\,S_{em}}{\hbar}\right]\int\,\mathcal{D}PF_1[\vett{P}]\int\,\mathcal{D}R\,F_2[\vett{P},\vett{R}],
\eeq
where $S_{em}=\int\,dt\,d^3x\,\mathcal{L}_{em}$ and the quantities $F_{1,2}$ are functionals of the matter and reservoir fields, defined as:
\barr
F_1[\vett{P}]&=&\exp\left[\frac{i}{\hbar}\int\,dt\,d^3x\left(\mathcal{L}_{matter}+\mathcal{L}_{lmi}\right)\right],\\
F_2[\vett{P},\vett{R}]&=&\exp\left[\frac{i}{\hbar}\int\,dt\,d^3x\,\mathcal{L}_{res}\right].\label{intF2}
\earr

\noindent We are interested in calculating $F_2[\vett{P},\vett{R}]$ explicitly. Intuitively, since $\mathcal{L}_{res}$ only depends on $\vett{R}^2$ and $\dot{\vett{R}}^2$ [see Eq. \eqref{eq3t}], we can use Gaussian integrals to compute $F_2[\vett{P},\vett{R}]$. For continuous fields, they are defined as follows: \cite{srednicki, feynmanHibbs}
\beq\label{gaussianPath}
\int\mathcal{D}\phi\exp\left[-\frac{1}{2}\left\langle\phi,\hat{A}\phi\right\rangle+\langle \psi,\phi\rangle\right]=\exp\left[\frac{1}{2}\langle\,\psi,\hat{A}^{-1}\psi\rangle\right],
\eeq
where $\psi$ is an auxiliary field (source term), $\hat{A}$ is a differential operator (typically well-behaved and invertible), $\hat{A}^{-1}$ is the propagator associated to the field $\phi$, i.e., the inverse of the differential operator $\hat{A}$, and $\langle\,\psi,\phi\rangle$ is a shorthand for
\beq
\langle\,\psi,\phi\rangle=\int\,d^nx\,\psi^*(x)\phi(x),
\eeq
with $n$ being the dimension of the space in which the fields $\psi(x)$ and $\phi(x)$ are defined. 

\noindent To calculate the integral in Eq. (\ref{intF2}) using Gaussian path integrals, we first need to bring its exponent into a quadratic form similar to that of the integral above. To do so, let us first integrate by part the matter-reservoir terms, so that we shift the time derivative from the reservoir field to the matter field as follows:
\barr
&-&\int_0^{\infty}\,d\omega\,\int\,dt\,f(\omega)\,\vett{P}(t)\cdot\dot{\vett{R}}(t) \\ \nonumber
&=&\int_0^{\infty}\,d\omega\,\int\,dt\,f(\omega)\,\dot{\vett{P}}(t)\cdot\vett{R}(t)\nonumber\\
&=&\langle\,\vett{B}(\tau)\cdot\vett{R}(\tau)\rangle,
\earr
where in the third line we introduced the $\langle\,\cdot\,\rangle$ notation with $d^nx=d^3x\,dt\,d\tau\,d\omega$ and we defined:
\beq
B(\tau)=-\frac{i}{\hbar}f(\omega)\,\dot{\vett{P}}(\tau)\,\delta(t-\tau),
\eeq
to write this term already in a form compatible with Eq. (\ref{gaussianPath}). Notice, moreover, that we have introduced an extra integration with respect to the variable $\tau$ (and a correspondent Dirac delta function) for later convenience.

\noindent To transform the rest of $\mathcal{L}_{res}$ into a quadratic form like the one in Eq. (\ref{gaussianPath}), we first use the identity ($\partial_t$ denotes time derivation)
\beq
(\partial_tR)^2=\partial_t(R\partial_tR)-R\partial_t^2R,
\eeq
and then integrate by parts once more, so that we are left with one term proportional to $R\partial_t^2R$ and another term proportional to $\omega^2R^2$. Putting everything together, and introducing the differential operator $\hat{A}$ as:
\beq\label{eq13t}
\hat{A}(t,\tau)=\frac{i}{2\hbar}\left(\partial_t^2+\omega^2\right)\delta(t-\tau),
\eeq
we get, for the exponent of Eq. (\ref{intF2})
\barr
\frac{i}{\hbar}\int\,dt\,d^3x\,\mathcal{L}_{res}=&-&\frac{1}{2}\langle\, R(\tau),\hat{A}(t,\tau)\,R(t)\rangle \\ \nonumber 
&-&\langle\,B(\tau)R(t)\rangle,
\earr
We can then use Eq. (\ref{gaussianPath}) to calculate the path integral with respect to the reservoir degrees of freedom appearing in Eq. (\ref{intF2}), obtaining:
\barr
\int\,\mathcal{D}R\,F_2[\vett{P},\vett{R}]=\exp\left[\frac{1}{2}\left(B,\hat{A}^{-1}B\right)\right].
\earr
If we now use the definition of $B(\tau)$ given above  and define the inverse of the operator $\hat{A}$ as the Green's function, i.e., $\hat{A}^{-1}=G(t-\tau,\vett{x})$ \cite{fueller}, we get:
\barr
\frac{1}{2}\left(B,\hat{A}^{-1}B\right)=-\frac{i}{\hbar}\Big[\int\,dt\,dx\,\int_0^{\infty}\,d\omega\,|f(\omega)|^2\,\vett{P}^2(t)\nonumber\\
+\int\,dt\,d\tau\,dx\,\vett{P}(t)\cdot G(t-\tau,x)\cdot\vett{P}(\tau)\Big].
\earr
The calculation of the second path integral, to eliminate the matter degrees of freedom, can be done using the same line of reasoning. The only complication is that the expression of the differential operator needed to define a quadratic form in the matter field $P(x,t)$ is more complicated than $\hat{A}$, as it results in an integro-differential operator, whose Green's function is, in general, very complicated to calculate. The reason for this is the presence, in the result above, of the quadratic form $\vett{P}(t)\cdot G(t-\tau)\cdot\vett{P}(\tau)$, which includes the effects of the reservoir in the Green's function for the matter field. Details on how to calculate explicitly this second integral, to arrive to the final result presented in Eq. \eqref{eq16t}, are given in Refs. \cite{bechler,PhysRevA.100.053845}
\section{Derivation of Eq. \eqref{eq36t}}\label{appendixC}
\noindent To derive Eq. \eqref{eq36t}, we first need to represent the interaction Lagrangian in frequency domain. We do this at the action level, instead, since the action $S_{int}=\int\,dt\,\mathcal{L}_{int}=\int\,(d\omega/2\pi)\,\tilde{\mathcal{L}}_{int}$ is left invariant by this transformation.

If we use the Fourier representation of the electric field 
\beq
E(t)=\frac{1}{\sqrt{2\pi}}\int\,d\omega\,E(\omega)\,\exp\left(i\omega\, t\right),
\eeq
we can transform the Kerr interaction Lagrangian as follows:
\barr\label{eqC2}
\int\,dt\,\mathcal{L}_{int}&=&\frac{3\varepsilon_0\chi^{(3)}}{2}\int\,dt\,\frac{d\omega}{(2\pi)^2}\,[d\Omega]_3\,E^*(\Omega)E^*(\Omega')\nonumber\\
&\times&E(\Omega'')E(\omega)\exp\left[i(\omega+\Omega''-\Omega'-\Omega)t\right]\nonumber\\
&=&\int\,\frac{d\omega}{2\pi}\,E(\omega)F(\omega),
\earr
where $[d\Omega]_3$ is a shorthand for $d\Omega\,d\Omega'\,d\Omega''$, and
\beq
F(\omega)=\frac{3\varepsilon_0\chi^{(3)}}{2}\int\,d\Omega\,d\Omega'\,E^*(\Omega)E^*(\Omega')E(\Omega+\Omega'-\omega).
\eeq
To go from the second to the third line in Eq. \eqref{eqC2} we have used the definition of the Dirac delta function \cite{fueller}
\beq
\int\,dt\,\exp\left[i(\omega-\Omega)t\right]=2\pi\,\delta(\omega-\Omega).
\eeq
Since $S_{int}=\int\,(d\omega/2\pi)\,\tilde{\mathcal{L}}_{int}$ we then have that the interaction Lagrangian in frequency domain reads:
\beq\label{eq34t}
\tilde{\mathcal{L}}_{int}=\frac{3\varepsilon_0\chi^{(3)}}{2}\,E(\omega)\int\,d\Omega\,d\Omega'\,E^*(\Omega)E^*(\Omega')E(\Omega+\Omega'-\omega).
\eeq
We then calculate the derivative of the interaction Lagrangian with respect to the field $E^*(\omega)$, using the identity \cite{srednicki}
\beq\label{eqC6}
\frac{\partial E(\alpha)}{\partial E(\beta)}=\delta(\alpha-\beta),
\eeq
to otain:
\beq
\frac{1}{E(\omega)}\frac{\partial\mathcal{L}_{int}}{\partial\,E^*(\omega)}=3\varepsilon_0\chi^{(3)}\int\,d\Omega\,|E(\Omega)|^2=\frac{3\chi^{(3)}}{2cn}I,
\eeq
where to get the last equality we used the definition of the time-averaged intensity of the electromagnetic field \cite{boyd2008nonlinear}. 
\section{Derivation of Eq. \eqref{npEps}}\label{appendixD}
\noindent To derive Eq. \eqref{npEps}, we first need to express the BI action in frequency domain, and then use the relation
\barr\label{eqD5}
\varepsilon_{BI}(\omega)&=&\frac{1}{\varepsilon_0 E(\omega)}\frac{\partial\tilde{\mathcal{L}}_{BI}}{\partial E^*(\omega)},
\earr
to calculate the dielectric function. This is in general a very complicated task, because there is no closed-form expression for the Fourier transform of the square root of a function. To overcome this problem, we start by representing the BI action with its power series expansion, i.e., the power series expansion for the function $\sqrt{1-x}$, as follows:
\beq
\int\,dt\,\mathcal{L}_{BI}=\sum_{n=1}^{\infty}\,c_n\,\int\,dt\,\left(\frac{\varepsilon_0|E(t)|^2}{b^2}\right)^n,
\eeq
and the explicit expression of the expansion coefficients is given by $c_n=(-1)^{n+1}\binom{1/2}{n}$. 

The idea is then to find a Fourier representation for the terms $\int\,dt\,|E(t)|^{2n}$ separately, then take the derivative with respect to $E^*(\omega)$ and after this re-sum the result into a closed-form expression. For $n=1$ this is trivial, since $\int\,dt\,|E(t)|^2=\int\,d\omega\,|E(\omega)|^2$ by virtue of the Parseval theorem. For $n\geq 2$, the integrals have the following general expression
\beq
\int\,dt\,|E(t)|^{2n}=\int\,\frac{d\omega}{2\pi}E(\omega)G_n(\omega),
\eeq
where $G_n(\omega)$ in general contains the convolution of $2n-1$ fields and can be written in the following form
\barr\label{eqD3}
G_n(\omega)&=&\int\,[d\Omega]_{2n-2}\underbrace{E^*(\Omega_1)\cdots E^*(\Omega_n)}_{n-\text{terms}}\nonumber\\
&\times&\underbrace{E(\Omega_{n+1})\cdots E(\Omega_{2n-2})}_{(n-2)-\text{terms}}E([\Omega]_{n,n-2}-\omega),
\earr
where, $[d\Omega]_{2n-2}=d\Omega_1 d\Omega_2\cdots d\Omega_{2n-2}$, and $[\Omega]_{n,n-2}=\Omega_1+\Omega_2+\cdots+\Omega_n-\Omega_{n+1}-\cdots -\Omega_{2n-2}$.
Recalling that the Fourier representation of the Lagrangian is given formally as $S=\int\,(d\omega/2\pi)\tilde{\mathcal{L}}$, the result above allows us to write the BI Lagrangian in frequency space in the following form:
\beq\label{eqD4}
\tilde{\mathcal{L}}_{BI}=E(\omega)\,\sum_{k=1}^{\infty}(-1)^{k+1}\binom{1/2}{k}G_n(\omega).
\eeq
We now need to calculate $\partial G_n(\omega)/\partial E^*(\omega)$. Using Eqs. \eqref{eqD3} and \eqref{eqC6}, we get:
\barr
\frac{\partial G_n(\omega)}{\partial E^*(\omega)}&=&n\,\int\,[d\Omega]_{2n-3}\underbrace{E^*(\Omega_1)\cdots E^*(\Omega_{n-1})}_{(n-1)-\text{terms}}\nonumber\\
&\times&\underbrace{E(\Omega_n)\cdots E(\Omega_{2n-3})}_{(n-2)-\text{terms}}E([\Omega]_{n-1,n-2}).
\earr
The next step to get to Eq. \eqref{npEps} is to write the result above in terms of intensity. To do this, we need to insert the inverse Fourier representation for the fields $E(\Omega_k)$. Since the integral above contains $2(n-1)$ fields, there will be $2(n-1)$ time integrals and, ultimately, $2(n-1)$ Dirac delta functions cancelling that many frequency integrals. We will then be left with only one time integral of the form $\int\,dt\,|E(t)|^{2n}$. To convince ourselves of this statement, let us consider explicitly the calculations for the first nontrivial case, i.e., $n=3$. We then have
\barr
\frac{\partial G_3(\omega)}{\partial E^*(\omega)}&=&\int\,[d\Omega]_3E^*(\Omega_1)E^*(\Omega_2)E(\Omega_3)E(\Omega_1+\Omega_2-\Omega_3)\nonumber\\
&=&\int\,[d\Omega]_3[dt]_4\mathcal{E}([t]_4)\mathcal{D}([t]_4,[\Omega]_3)\nonumber\\
&=&\int\,[dt]_4\mathcal{E}([t]_4)\delta(t_1-t_4)\delta(t_2-t_4)\delta(t_3-t_4)\nonumber\\
&=&\int\,dt\,|E(t)|^4,
\earr
where in the second line we have defined
\begin{subequations}
\begin{align}
\mathcal{E}([t]_4)&=E^*(t_1)E^*(t_2)E(t_3)E(t_4),\\   
\mathcal{D}([t]_4,[\Omega]_3)&=\exp\Bigg[i\Bigg(\Omega_1t_1+\Omega_2t_2\nonumber\\
&-\Omega_3t_3-(\Omega_1+\Omega_2-\Omega_3)t_4\Bigg)\Bigg].
\end{align}
\end{subequations}
Generalising this procedure then leads to the following result:
\beq
\frac{\partial G_n}{\partial E^*(\omega)}=n\int\,dt\,|E(t)|^{2n}.
\eeq
Substituting this result into Eq. \eqref{eqD5}, using the identity
\beq
\sum_{n=1}^{\infty}(-1)^{n+1}\binom{1/2}{n}n\,x^n=\frac{x}{2\sqrt{1-x}},
\eeq
and using the relation between time-averaged electric field and intensity leads then to Eq. \eqref{npEps}.
\bibliography{sn-bibliography}

\end{document}